\documentclass[a4paper,11pt]{article}
\usepackage{jheppub} 
\usepackage{lineno}
\usepackage[colorinlistoftodos]{todonotes}
\usepackage{appendix}
\usepackage{amsthm}
\usepackage{bm}
\usepackage{braket}
\usepackage{amsmath}
\usepackage{amsfonts}
\usepackage{mathtools}
\usepackage{graphicx}
\usepackage{cancel}
\usepackage{slashed}
\usepackage{wrapfig}
\usepackage{nextpage}
\usepackage{enumitem}
\usepackage{float}
\usepackage{hyperref}
\usepackage{amssymb}
\usepackage{enumitem}
\usepackage{simpler-wick}
\hypersetup{colorlinks=true, urlcolor=black, citecolor=blue, linkcolor=black}

\usepackage{graphicx} 
\usepackage{nicematrix}

\newcommand{\Tr}{\text{Tr}}

\title{\boldmath A Renormalization Group Analysis of the Ising Model Coupled to Causal Dynamical Triangulations}

\author[1]{R. Barouki,\note{Corresponding authors.}}
\author[1]{and D. Laurenzano}

\affiliation{Rudolf Peierls Centre for Theoretical Physics, University of Oxford \\ Parks Road, Oxford OX1 3PU, United Kingdom}

\emailAdd{ryan.barouki@physics.ox.ac.uk}
\emailAdd{davide.laurenzano@physics.ox.ac.uk}

\abstract{
We analyze the matrix model characterizing the Ising model coupled to Causal Dynamical Triangulations (CDT) from the point of view of the Functional Renormalization Group Equation (FRGE). This model is a dually weighted matrix model, whose Feynman diagrams are in correspondence with discrete triangulations of two-dimensional geometries with a preferred time foliation. In particular, we find the fixed points of the beta-function equations, showing that the number of relevant directions in each case is compatible with the physical interpretation of the CFT living on the fixed points. In addition to recovering the fixed points for topological gravity and pure gravity with a cosmological constant, we find a new fixed point featuring three relevant directions which match the number of primary fields in the Ising CFT.
}

\begin{document}

\maketitle
\section{Introduction}
The problem of finding a consistent theory of quantum gravity has long been a central challenge in theoretical physics. In this paper, we focus on \textbf{two-dimensional quantum gravity}, which can be understood as a theory of random surfaces. An effective way to tackle random surfaces is to discretize them, typically by \textit{triangulating}—i.e.\ requiring that all faces of the embedded graph be triangular. The functional integral over the metrics then becomes a sum over random surfaces
$$
\int_{\mathcal M}\mathcal Dg~~\rightarrow \sum_{\text{triangulations of }\mathcal M}
$$
and the Einstein-Hilbert action is also discretized. This was first done by Regge \cite{Regge1961} for numerical General Relativity. This discretization introduces a regulator in the form of the triangle edge lengths, making computations more tractable, after which one can take a continuum limit to recover the original continuous geometry.

An ingenious approach to generate all possible triangulations of a surface was first proposed by 't Hooft in
\cite{Thooft1974} wherein he introduced the concept of matrix models. He showed that one can use these simple matrix
integrals to generate all possible triangulations by essentially generating ``fat" Feynman diagrams, where the ``fatness" of the graphs creates internal loops that can be thought of as faces. Thus, whereas Gaussian integrals of vector
or scalar quantities enumerate objects with vertices and edges (graphs), Gaussian integrals of matrices enumerate
objects with vertices, edges and faces (maps/embedded graphs). They have since been used extensively to study many aspects of two-dimensional quantum gravity with some notable recent examples \cite{Saad2019, Collier, Collier2024}. See \cite{DiFrancesco1993} for an older review and \cite{Anninos2020} for a more recent one.

When summing over a random discrete surface, there is some freedom in deciding which ensemble of graphs to include in the sum. One choice is the ensemble of planar graphs, often called the Euclidean ensemble. While this leads to many interesting results consistent with the continuum perspective of Liouville quantum gravity, it also encounters serious issues. One prominent example is the unexpected value of the Hausdorff dimension, $d_H=4$, which seems to contradict the naive expectation of $d_H=2$. These difficulties motivated the development of the \textbf{Causal Dynamical Triangulations (CDT)} program, which imposes a causal structure on the graph ensemble by demanding a global time foliation \cite{Ambjorn1998}.

In CDT, edges are split into two types—time-like and space-like—ensuring a clear causal structure. This distinction allows for well-defined Wick rotations so that one can work in Euclidean signature while still preserving essential Lorentzian features such as causality. CDT has been remarkably successful in the non-perturbative study of two-dimensional quantum gravity, with many analytic and numerical results \cite{Durhuus2009sm, Ambj_rn_2006, Ambj_rn_1999, Ambj_rn_2000, Ambj_rn_2009, ambjorn_c_2015, Ambj_rn_2013, Wheater_2022, durhuus2022trees, Barouki2025}. Moreover, considerable progress has been made in four-dimensional CDT using numerical methods \cite{Ambjorn2005, Loll2005, Ambjorn2007}.

A description of quantum gravity would not be complete without understanding the interplay between matter and geometry. To that end, we study
two-dimensional CDT coupled to a simple matter model: the Ising model. Specifically, we adopt a matrix model
representation of this coupled Ising–CDT system, as introduced in \cite{Abranches}, and investigate its fixed-point
structure using Functional Renormalization Group (FRG) methods. Fixed points of the FRG represent potential double scaling limits of the matrix model, where $N\rightarrow \infty$ and couplings scale towards a critical value, the scaling being characterized by a critical exponent. Following the approach of \cite{Eichhorn2013}, where FRG
techniques were applied to tensor models, we extend the results of Castro and Koslowski
\cite{castro2020renormalization}, who first applied FRG methods to the pure CDT matrix model by Benedetti and Henson
\cite{Benedetti}. In particular, we demonstrate the existence of fixed points of RG flow in the Ising–CDT matrix
model with precisely the right number of relevant directions to describe a continuum Ising CFT. More specifically, we
show that the number of relevant directions matches the three primary fields of the Ising CFT \cite{Boulatov1986}. Moreover, we find other types of fixed points with different numbers of relevant directions. These could, in principle, represent different continuum limits of the discrete theory. One of our main findings is that the parameter space has a rich structure, including segments such that all points on the segment are fixed points of the RG flow.

The paper is organized as follows. In Section \ref{sec:MatrixIsing}, we review the main features of the pure CDT and Ising--CDT matrix models. In Section \ref{sec:FRGE}, we set up the Functional Renormalization Group Equation (FRGE) for the Ising–CDT matrix model. Section \ref{sec:Beta} presents the derivation of the general form of the beta-function equations. Finally, we analyze the fixed-point structure in Section \ref{sec:Fixed}.

\section{A Matrix Model for Ising CDT}
\label{sec:MatrixIsing}

\begin{figure}
  \begin{center}
    \includegraphics[width=0.85\textwidth]{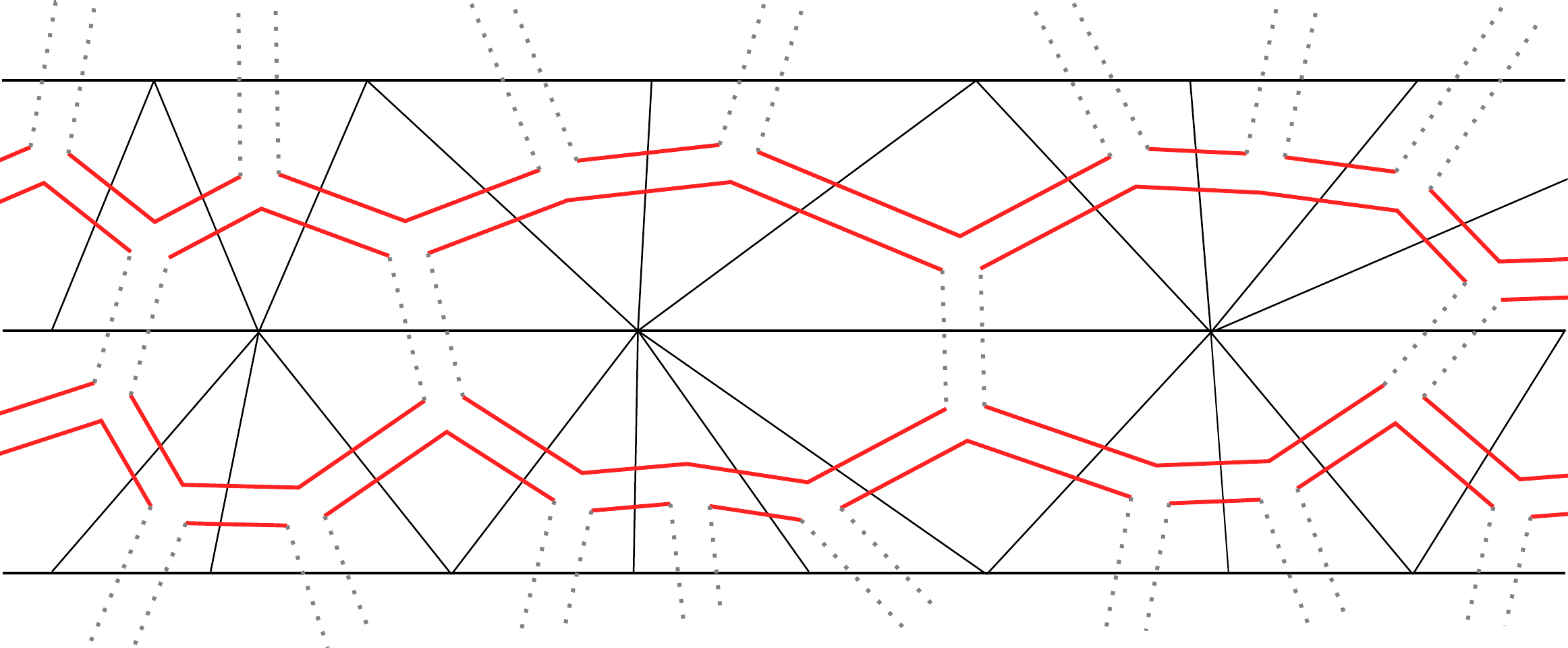}
  \end{center}
  \caption{An example of a CDT strip and the dual ribbon graph. The red lines are the propagators for the $A$ matrix and the gray dashed lines are the $B$ propagators.}\label{fig:CDTRibbon}
\end{figure}

The matrix model for pure CDT was introduced by Benedetti and Henson (BH) \cite{Benedetti} and is defined by the following partition function
\begin{equation*}
    Z=\int \mathcal{D}A \, \mathcal{D}B e^{-N\Tr[\frac{1}{2}A^2+\frac{1}{2}(C^{-1}B)^2-gA^2B]} \, ,
\end{equation*}
with Hermitian matrices $A$ and $B$, along with the constant matrix $C$ that satisfies the condition

\begin{equation}
    \Tr[C^m]=N \delta_{m,2} \, .
    \label{eq:ConditionC}
\end{equation}
The propagators are given by
\begin{equation*}
    \langle A_{ij} A_{kl} \rangle = \frac{1}{N} \delta_{i l} \delta_{j k}, \quad \langle B_{ij} B_{kl} \rangle = \frac{1}{N} C_{il} C_{j k}, \quad \langle A_{ij} B_{kl} \rangle =0 \, .
\end{equation*}
The two matrices $A$ and $B$ are necessary to capture the two different types of edges in a dual CDT graph: space-like edges represented by the matrix $A$ and time-like edges represented by the matrix $B$. Note that time-like edges in the dual graph correspond to space-like edges in the triangulation and vice-versa.
The structure of the Feynman graphs generated by this model consists of $AAB$ vertices and faces with two time-like edges, with the exception of the initial and final slices of the graph where the faces have 0 time-like edges (see Figure \ref{fig:CDTRibbon}). Each face with $m$ internal $B$ lines contributes a factor of $\Tr[C^m]$, hence the condition (\ref{eq:ConditionC}) ensures that each face has either two or zero $B$-lines (time-like edges). These local requirements are sufficient to generate valid CDT graphs with a global foliation \cite{Benedetti}.

The $B$ field can be integrated out to obtain a simpler expression only in terms of the $A$ matrix:
\begin{equation}
  Z=\int \mathcal{D}A\, e^{-N\Tr[\frac{1}{2}A^2-\frac{g^2}{2}(A^2C)^2]} \, .
    \label{eq:PureCDTMatrix}
\end{equation}
The matrix model (\ref{eq:PureCDTMatrix}) was considered in \cite{castro2020renormalization} from the point of view of the Functional Renormalization Group Equation (FRGE). In particular, it was shown that the theory admits different possible continuum limits, represented by the UV fixed points of the Renormalization Group flow obtained by varying the matrix size $N$. 

This work extends the analysis to CDT coupled to the Ising model. In order to do so, we need a matrix model which describes the coupling of the Ising model to CDT. In particular, the theory must account for the spin degree of freedom at each vertex of the dual fat graph, which takes values in $\{-1, +1\}$. Moreover, the edges connecting vertices with the same or different spins must be distinguished. The matrix model that accounts for these requirements is the four-matrix model given by \cite{Abranches}, which generalises the model by BH:
\begin{equation*}
    Z=\int \mathcal{D}A_+ \mathcal{D}A_- \mathcal{D}B_+ \mathcal{D}B_-\,e^{-NS} \, ,
\end{equation*}
where the action $S$ is
\begin{multline*}
    S=\Tr\Bigl[\frac{1}{2}A_+^2+\frac{1}{2}(C^{-1}B_+)^2+\frac{1}{2}A_-^2+\frac{1}{2}(C^{-1}B_-)^2\\
    -\gamma A_+ A_- -\gamma (C^{-1}B_+)(C^{-1}B_-)-gA_+^2B_+-gA_-^2B_-\Bigr] \,.
\end{multline*}
The matrix $C$ satisfies the property (\ref{eq:ConditionC}) and the propagators are given by
\begin{align*}
    \langle A_{+,ij} A_{+,kl} \rangle &=  \langle A_{-,ij} A_{-,kl} \rangle =\frac{1}{N} \frac{1}{1-\gamma^2}\delta_{i l} \delta_{j k}, \\
    \langle B_{+,ij} B_{+,kl} \rangle &= \langle B_{-,ij} B_{-,kl} \rangle=  \frac{1}{N} \frac{1}{1-\gamma^2}C_{il} C_{j k}, \\
    \langle A_{+,ij} A_{-,kl} \rangle &=\frac{1}{N} \frac{\gamma}{1-\gamma^2}\delta_{i l} \delta_{j k}, \\
    \langle B_{+,ij} B_{-,kl} \rangle &=\frac{1}{N} \frac{\gamma}{1-\gamma^2}C_{il} C_{j k}, \quad \langle A_{I,ij}B_{J, kl} \rangle=0,
\end{align*}
where $I,J=+,-$ in the last equation. Once again, the matrices $B_+$ and $B_-$ can be integrated out to obtain a simplified expression
\begin{equation}
Z=\int \mathcal{D}A_+ \mathcal{D}A_-\,
e^{-N\Tr\Bigl[\frac{1}{2}A_+^2+\frac{1}{2}A_-^2-\gamma A_+ A_- -\frac{1}{2}\frac{g^2}{1-\gamma}(A_+^2C)^2-\frac{1}{2}\frac{g^2}{1-\gamma}(A_-^2C)^2-\frac{1}{2}\frac{\gamma g^2}{1-\gamma^2}\bigl((A_+^2+A_-^2)C\bigr)^2\Bigr]}  \, .
    \label{eq:IsingCDT}
\end{equation}
It is convenient to define $U = (A_+ + A_-)/\sqrt  2$ and $V = (A_+ - A_-)/\sqrt  2$ in order to diagonalize the kinetic part of the action.
\begin{equation}
Z=\int \mathcal{D}U \mathcal{D}V\,
e^{-N\Tr\Bigl[\frac{1}{2}(1-\gamma)U^2+\frac{1}{2}(1+\gamma)V^2 -\frac{1}{4}\frac{g^2}{1-\gamma}((U^2 + V^2)C)^2-\frac{1}{4}\frac{g^2}{1+\gamma}((UV + VU)C)^2\Bigr]}  \, .
    \label{eq:IsingCDTDiag}
\end{equation}
This transformation takes the Ising model with spins defined on the vertices of the ribbon graph to one where the spins live on the faces. Since the
triangulation is the dual of the ribbon graph, the Ising spins now live on the vertices of the CDT. The matrices now have the
interpretation that $U$ defines an edge in the ribbon graph that separates faces of equal parity (i.e. both spin up or both spin down) and $V$ separates
faces of opposite parity. The action in (\ref{eq:IsingCDTDiag}) is symmetric under $U \to -U$ and $V \to -V$ and hence has a $\mathbb Z_2 \times \mathbb Z_2$ symmetry. This symmetry implements the logical consistency condition that the
four spins that live on the faces around the 4-valent vertices cannot exist in any configuration where only one edge has
opposite parity and the other three are all the same. The same is true if three of the edges have opposite parity; then
this implies the fourth edge must have opposite parity. Hence, $U$s and $V$s can only come in pairs.
Furthermore, we follow \cite{castro2020renormalization} and write the action as 
\begin{multline}
  S = N\Tr\Bigl[\frac{1}{2}(1-\gamma)UU^T+\frac{1}{2}(1+\gamma)VV^T \\
  -\frac{1}{4}\frac{g^2}{1-\gamma}\bigl((UU^T + VV^T)C\bigr)^2-\frac{1}{4}\frac{g^2}{1+\gamma}\bigl((UV^T + VU^T)C\bigr)^2\Bigr],
\label{eq:IsingCDTActionUV}
\end{multline}
This creates a symmetry in the index structure of the action, which is convenient in our calculations. It does not change any of the above discussion and, from now on, we will deal exclusively with this action.

Analyzing this matrix model from a Functional Renormalization Group perspective will be the goal of the rest of the paper. In particular, we will show that it admits a continuum limit at fixed points of the FRGE flow.
\section{Functional Renormalization Group Equation}
\label{sec:FRGE}
In this section, we discuss the Functional Renormalization Group Equation (FRGE) for the Ising-CDT matrix model given in (\ref{eq:IsingCDTActionUV}), which will allow us to derive the $\beta$-function equations in the following section. 

The following sections involve heavy use of index notation. To help with readability, we summarise our conventions in the following table.
\begin{table}[H]
\centering
\begin{tabular}{|l|l|}
\hline
Symbol & Meaning \\
\hline
$a,b,c,d$ & Matrix indices in $\{1,\dots,N\}$ \\
$I,J,K\in\{1,2\}$ & ``Parity'' indices: $U_I \in \{U_1, U_2\}$ where $U_1:= U$, $U_2:= V$ \\
$i,j,k,l,h,m$ & Labels for couplings $g_4^{(i)}, g_6^{(h)}$ of distinct operators $O_4^{(i)}, O_6^{(h)}$ \\
\hline
\end{tabular}
\caption{Index and symbol conventions used in Sections~\ref{sec:FRGE} and \ref{sec:Beta}.}
\label{tab:notation}
\end{table}

The FRGE is a differential equation for a momentum-dependent effective average action $\Gamma_k[\phi]$. See \cite{Wetterich_1993} for a general treatment and \cite{Eichhorn2013} for an application to matrix models. We provide a brief introduction as it applies to (\ref{eq:IsingCDT}) here.

The idea is to construct an effective action where momentum modes above the scale $k$ have been integrated out while maintaining those modes below $k$. The trick introduced in \cite{Wetterich_1993} is to give a momentum-dependent mass to modes below the scale $k$, which suppresses them in the functional integral and allows only the higher modes to be integrated.
This is achieved by adding a term $\Delta S_k=\frac{1}{2}\phi R_k \phi$ to the action. 
We can then define the effective generating functional $W_k[J]$ as
$$
e^{-W_k[J]} = \int \mathcal{D}\phi e^{-S[\phi] + J\cdot\phi - \Delta S_k},
$$
and the effective action as a modified Legendre transform 
\begin{equation}
\label{eq:gamma_k}
\Gamma_k[\phi] = \sup_J\{\int d^n x J(x)\phi(x) + W_k[J]\} - \Delta S_k.
\end{equation}
Therefore, $\Gamma_k[\phi]$ interpolates between the quantum 1PI effective action and the bare action:
$\lim_{k\to 0}\Gamma_k=\Gamma$ and $\lim_{k\to\Lambda}\Gamma_k=S$.
The IR-suppression term $R_k$ gives modes with $p^2<k^2$ an effective mass of order $k$
(heuristically, $R_k(p)\sim k^2\,\theta(k^2-p^2)$), which suppresses their fluctuations in the path integral.
In the matrix-model realization used here, the role of $k$ is played by the matrix size $N$ and
$R_N$ acts as a mass term of order $N$ for index modes with $(a+b)/2 < N$ [cf. eq.~\eqref{eq:IRsuppression}].

The form of $R_k(p)$ in momentum space can be anything that satisfies the following conditions. It must give a mass to modes $p^2<k^2$ and smoothly decrease to zero for $k^2<p^2<\Lambda^2$, where $\Lambda$ is a UV cutoff. It must also satisfy, 
$$
\lim_{k\to0} R_k(p) = 0
$$
which ensures that 
$$
\lim_{k\to0} \Gamma_k[\phi] = \Gamma[\phi],
$$
the full effective action for the 1PI correlation functions.
Finally, $R_k(p)$ must satisfy
$$
\lim_{k\to\Lambda\to\infty} R_k(p) = \infty,
$$
which ensures the saddle-point evaluation of the functional integral so that
$$
\lim_{k\to\Lambda\to\infty} \Gamma_k[\phi] = S[\phi].
$$
In the case of tensor (and matrix) models, $\phi_{a_1,..., a_m}$ is a rank-$m$ tensor and the scale $k$ is given by the tensor (matrix) size $N$.
The suppression term essentially gives a mass to the matrix entries $a,b = 1,...,N$. 
For the model in (\ref{eq:IsingCDT}), we define the effective generating functional as
\begin{equation*}
  e^{-W_N[J]}=\frac{1}{\mathcal{N}_N} \int_\Lambda\mathcal{D}U \mathcal{D}V e^{-S[U, V]+J_{ab}^U U^{ab} +J_{ab}^V V^{ab} - \Delta S_N[U,V]}\, .
\end{equation*}
where $\mathcal{N}_N$ is an $N$-dependent normalization factor, $\Lambda$ is a UV cutoff and $\Delta S_N[U,V]$ is the IR suppression given by
\begin{equation*}
  \Delta S_N[U,V]=\sum_{I,J = 1}^2\frac{1}{2}U_I^{ab}R_{N,abcd}^{IJ} U_J^{cd} \,,
\end{equation*}
where we define $U_1 := U, U_2 := V$ for notational convenience. 
Then by analogy to (\ref{eq:gamma_k}), the effective action is given by
\begin{equation}
    \Gamma_N[U,V]=\sup_J\{J^U_{ab}U^{ab} + J^V_{ab}V^{ab} + W_N[J]\} - \Delta S_N[U,V] \,.
    \label{eq:EffectiveAction}
\end{equation}
In \cite{Wetterich_1993}, Wetterich derived the following equation for $\Gamma_N$ under a change of the IR scale $N$
\begin{equation}
    \partial_{t}\Gamma_N=\frac{1}{2}\Tr\Bigl(\frac{\partial_t R_N}{R_N+\Gamma^{(2)}_N}\Bigr)\, ,
    \label{eq:FRGE}
\end{equation}
where $t=\log N$ and $\Gamma^{(2)}_N$ is the second derivative with respect to the matrix components 
\begin{equation*}
  \Gamma^{(2)\,}_{N,\,abcd}[U,V]_{IJ}=\frac{\delta}{\delta U^{ab}_I} \frac{\delta}{\delta U^{cd}_J}\Gamma_N[U,V] \, .
\end{equation*}
Equation (\ref{eq:FRGE}) is the Functional Renormalization Group Equation (FRGE).

The usual procedure to derive the effective action consists of writing down the most generic action compatible with symmetries. Then a suitable truncation is chosen to make explicit computations possible. Equation (\ref{eq:FRGE}) translates into a beta-function equations for the coupling constants of operators in the effective action. As we explained in the previous section, we consider all operators with an even number of $U$s and $V$s. Furthermore, we have an $O(N)$ symmetry acting on the right of the matrix fields
\begin{equation}
  U \to UO,~~~V \to VO.
  \label{eq:ONsymmetry}
\end{equation}
Therefore, we only consider the combinations $UU^T$, $UV^T$, $VU^T$ and $VV^T$ in the effective action.

Following \cite{castro2020renormalization}, we choose a truncation of the effective action that only includes operators with two $C$ matrices. We also restrict to the set of single trace operators, as this was shown to be a good approximation for matrix models \cite{castro2020renormalization, Eichhorn2013}.

With this in mind, we can write the effective action as 
\begin{equation}
  \Gamma_N = \frac{Z_U}{2} \,\Tr\Bigl[UU^T\Bigr] + \frac{Z_V}{2} \,\Tr\Bigl[VV^T\Bigr]+\sum_{n=2}^{\infty}\, \,\sum_{j}\frac{\Bar{g}_{2n}^{(j)}}{2n}O_{2n}^{(j)} \, ,
  \label{eq:EffectiveActionTruncation}
\end{equation}
where $Z_U$ and $Z_V$ are wavefunction renormalization factors.
The operators $O_{2n}^{(j)}$ are defined as
\begin{equation}
  \begin{aligned}
    O_{4}^{(j)} &= \Tr\Bigl[U_{I_1} U_{I_2}^TC U_{I_3} U_{I_4}^T C\Bigr] \\
    O_{2n}^{(j)} &= \Tr\Bigl[U_{I_1} U_{I_2}^TC U_{I_3} U_{I_4}^T C\underbrace{U_{I_5} U_{I_6}^T ... U_{I_{2n-1}} U_{I_{2n}}^T}_{n-2~\text{terms}}\Bigr],~~n>2,\\ 
  \end{aligned}
  \label{eq:OperatorDefs}
\end{equation}
and $I_k \in \{1,2\}, \forall k$, $U_1 = U, U_2 = V$ as before and $j$ labels the set of configurations permitted by the
symmetries of the action discussed above.
The ones that we will consider explicitly are $O_4^{(j)}$ and 
\begin{equation*}
    O_{6}^{(j)} = \Tr\Bigl[U_{I_1} U_{I_2}^TC U_{I_3} U_{I_4}^T CU_{I_5} U_{I_6}^T\Bigr].
\end{equation*}
Following \cite{castro2020renormalization}, we truncate the expansion in (\ref{eq:EffectiveActionTruncation}) at $n=3$, to account for the terms which contribute to the beta function of the bare action at one loop. 
There are $5$ distinct $O_4$ operators allowed by the symmetries so $j = 1, ..., 5$. The symmetry allows any operator that
has an even number of $U$s and $V$s so we count them by simply summing the even binomial coefficients $\sum_{k=0}^2 {4\choose2k} = 8$. However,
this over-counts because
\begin{equation*}
    \Tr[UU^TCVV^TC]=\Tr[VV^TCUU^TC]
\end{equation*}
due to the cyclic property of the trace. The same is true for
\begin{equation*}
    \Tr[UV^TCVU^TC]=\Tr[VU^TCUV^TC] \, .
\end{equation*}
Moreover 
\begin{equation*}
    UV^TCUV^TC=(VU^TCVU^TC)^T \, ,
\end{equation*}
implying that their trace yields the same operator, namely
\begin{equation*}
    \Tr[UV^TCUV^TC]=\Tr[VU^TCVU^TC] \, .
\end{equation*}
Hence, we are left with 5 distinct operators at fourth-order. 
For $O_6$ operators, the over-counting is not due to the cyclicity of the trace because, in our basis, the two $C$ matrices
sit at fixed positions. Cyclic permutations of the index pattern generally move the $C$ insertions and therefore produce distinct basis elements within our truncation.
After modding out only by reversal (transpose) symmetry, the $2^5=32$ naive patterns reduce to $20$ distinct $O_6$ operators.

We introduce the dimensionless couplings $g_{2n}^{(j)}$ for $n=2,3$ after re-scaling the kinetic term to its canonical coefficient, defined by
\begin{equation}
  \Bar{g}_{4}^{(j)}=Z_U^{N_U^{(j)}/2} Z_V^{N_V^{(j)}/2}N^{\alpha_4} g_4^{(j)}, \quad \Bar{g}_{6}^{(j)}=Z_U^{N_U^{(j)}/2} Z_V^{N_V^{(j)}/2} N^{\alpha_6} g_6^{(j)}\,,
  \label{eq:CanonicalScalings}
\end{equation}
where $N_U^{(j)}$ and $N_V^{(j)}$ are the numbers of $U$ and $V$ matrices respectively in the operator. For the $O_4$
operators, $N_U^{(j)} + N_V^{(j)} = 4$ and $N_U^{(j)} + N_V^{(j)} = 6$ for $O_6$ operators.
The canonical scaling dimensions $\alpha_4$ and $\alpha_6$ are not determined by dimensional analysis because the RG flow scale is the dimensionless matrix size $N$. However, they can be fixed by requiring that the beta functions have no explicit dependence on $N$ (they are ``autonomous'') in the large $N$ limit. This sets an upper bound on the canonical scaling of the couplings \cite{Eichhorn2018}. Choosing the canonical dimensions below this bound only removes interactions from the beta functions, so the ``most interacting" theory is one where the bound is saturated.

Having specified a consistent truncation of the effective action (\ref{eq:EffectiveActionTruncation}), we can now proceed to analyse the FRGE (\ref{eq:FRGE}) in more detail. First, we define $G_N^{(2)}$ as the term in $\Gamma_N^{(2)}$ which does not depend on any $U,V$ field. Given equation (\ref{eq:EffectiveActionTruncation}), this is given by
\begin{equation}
    G^{(2),abcd}_{N}= \begin{pmatrix}
        Z_U & 0 \\
        0 & Z_V
    \end{pmatrix}
    \delta^{ac} \delta^{bd} \, .
    \label{eq:EffectiveActionFieldIndep}
\end{equation}
With this split, we can consider the expansion of the right-hand side of the FRGE (\ref{eq:FRGE})
\begin{equation}
    \frac{1}{2}\Tr\Bigl(\frac{\partial_t R_N}{R_N+\Gamma^{(2)}_N}\Bigr)=\frac{1}{2}\Tr\Bigl(\Dot{R}P\Bigr)-\frac{1}{2}\Tr\Bigl(\Dot{R}PFP\Bigr)+ \frac{1}{2}\Tr\Bigl(\Dot{R}PFPFP\Bigr)+...
    \label{eq:FRGEexpansion}
\end{equation}
where $\Dot{R}$ indicates the derivative of $R$ with respect to $t=\log N$, 
\begin{equation*}
    P_N=(R_N+G_N^{(2)})^{-1}
\end{equation*}
is a field-independent term and $F$ is the field-dependent term in $\Gamma_N^{(2)}$. Explicitly, for the truncation of (\ref{eq:EffectiveActionTruncation}) to $n=3$, 
\begin{equation}
  F=\sum_{j=1}^6\Bar{g}_4^{(j)}F_{N, (j)}^{(4)}+\sum_{j'=1}^{32}\Bar{g}_6^{(j')}F_{N, (j')}^{(6)}
    \label{eq:FieldsOperator}
\end{equation}
$F_{N, (j)}^{(4)}$ and $F_{N, (j)}^{(6)}$ are the terms obtained by taking the second derivative with respect to the matrix elements of operators containing four and six fields respectively, namely
\begin{align*}
  F_{N, (j), PQ}^{(4),abcd}&=\frac{\delta}{\delta U_P^{ab}}\frac{\delta}{\delta U_Q^{cd}} O_4^{(j)}\\
  F_{N, (j), PQ}^{(6),abcd}&=\frac{\delta}{\delta U_P^{ab}}\frac{\delta}{\delta U_Q^{cd}} O_6^{(j)}
\end{align*}
Given the structure of the expansion (\ref{eq:FRGEexpansion}) and the definition of the propagator $P$, we generalize the definition given in \cite{castro2020renormalization} of the IR-suppression term $R_N$ to
\begin{equation}
    R_{N, IJ}^{abcd} := \Bigl( \frac{N}{a+b}-1 \Bigr) \theta\Bigl[1-\frac{a+b}{N}\Bigr]\begin{pmatrix}
        Z_U & 0 \\
        0 & Z_V
    \end{pmatrix}_{IJ}\delta^{ac}\delta^{bd} \, ,
    \label{eq:IRsuppression}
\end{equation}
so that the sign indices factorize in $G_N=R_N+F_N^{(2)}$ and we can invert it separately. It follows that
\begin{equation*}
    P^{abcd}_{N, IJ}=P_N^{abcd} \otimes P^{\text{sign}}_{IJ}
\end{equation*}
where 
\begin{equation*}
    P^{\text{sign}}_{IJ}=\begin{pmatrix}
      Z_U^{-1} & 0 \\
      0 & Z_V^{-1}
    \end{pmatrix}_{IJ}
\end{equation*}
is obtained by inverting the sign matrix in (\ref{eq:EffectiveActionFieldIndep}) and (\ref{eq:IRsuppression}). Note that the propagators and the IR cutoff term in (\ref{eq:FRGEexpansion}) are proportional to the identity in matrix space. Therefore, they will just contribute an overall $N$ dependent factor, which we will keep track of in the final result. For the sake of notation, we will only keep track of the $P^{\text{sign}}$ part of the propagator and suppress the dependence on $\Dot{R}$ and $P_N$ in the intermediate steps of the computation.   
\section{Beta Function Equations}
\label{sec:Beta}
Here we derive the beta functions within our single-trace, two-$C$ truncation.
We first extract the tensor structures that can appear on the right-hand side of the FRGE, then project them onto the
operator basis $O_4^{(i)}$ and $O_6^{(j)}$, making the $N$-scaling explicit.

\subsection{General form of the beta function equations}
To find the structure of the beta functions, it is sufficient to only consider the terms $F$ and $P^{\text{sign}}$ in \eqref{eq:FRGEexpansion} -- this
is what we will do in this section. In the next subsection, we will be more careful in order to deduce the $N$ dependence of the beta function equations. 
Let us first consider the left-hand side of the FRGE (\ref{eq:FRGE}). The derivative acts on the couplings in (\ref{eq:EffectiveActionTruncation}), resulting in 
\begin{equation}
  \begin{aligned}
    \partial_{t}\Gamma_N&=\frac{Z_U}{2}\eta_U\Tr\big[UU^T\big] + \frac{Z_V}{2}\eta_V \,\Tr\big[VV^T\big] 
+\sum_{n=2}^{\infty}\, \,\sum_{j}\frac{\Bar{\beta}_{2n}^{(j)}}{2n}O_{2n}^{(j)} \, \\
    &= \frac{Z_U}{2}\eta_U\Tr\big[UU^T\big] + \frac{Z_V}{2}\eta_V \,\Tr\big[VV^T\big] \\
    &+\sum_{n=2}^{\infty}\, \,\sum_{j}\frac{Z_U^{N_U^{(j)}/2}Z_V^{N_V^{(j)}/2}N^{\alpha_{2n}}(\beta_{2n}^{(j)}
    + \eta_U N_U^{(j)}/2 +\eta_V N_V^{(j)}/2 + \alpha_{2n} g_{2n}^{(j)})}{2n}O_{2n}^{(j)} \, \\
     \label{eq:LHSFRGE}
   \end{aligned}
\end{equation}
where $\eta_{U/V}=\partial_t \log Z_{U/V}$ is the anomalous dimension of the operators $\Tr\big[UU^T\big]$ and
$\Tr\big[VV^T\big]$ and 
\begin{equation*}
    \Bar{\beta}_4^{(j)}=\partial_t \Bar{g}_4^{(j)}, \qquad \Bar{\beta}_6^{(k)}=\partial_t\Bar{g}_6^{(k)},
\end{equation*}
are the beta functions for $\Bar{g}_4$ and $\Bar{g}_6$ respectively. In the last line, we used the definitions of the dimensionless couplings in (\ref{eq:CanonicalScalings}).

The next step is to compare this with the various terms on the right-hand side of the FRGE, namely the expansion (\ref{eq:FRGEexpansion}), looking for operators which match those in our chosen truncation (\ref{eq:LHSFRGE}). This defines the projection rule onto a smaller set of operators, and by comparing the pre-factors, we can read off the beta functions. 
\paragraph{Projection rule.}
Let $F_N[U,V]$ be the right-hand side of the FRGE expanded to the order of fields relevant for a given beta function.
We project onto a basis operator $O_{2n}^{(j)}$ by:
\begin{enumerate}[label=(\alph*),nosep]
\item discarding multi-trace terms (our truncation is single-trace);
\item discarding monomials with $C$-power $\neq 2$ inside a single trace;
\item matching the sequence of $U,V$ and transposes to the canonical pattern of $O_{2n}^{(j)}$ (up to overall cyclic rotation
and reversal, cf. the discussion above).
\end{enumerate}
The coefficient left in front of $O_{2n}^{(j)}$ after these steps defines the projected contribution to $\beta_{2n}^{(j)}$.

Let us consider the $F^{(4)}_N$ contribution from \eqref{eq:FieldsOperator} to first order in \eqref{eq:FRGEexpansion} given by:
\begin{equation}
  \sum_{j}\Bar{g}_4^{(j)} \sum_{\{I\}}\Bigl(c^{\{I\}}_{(j)}\Tr[C]\Tr[U^T_{I_1} C U_{I_2}]+(d^{\{I\}}_{(j)}+N f^{\{I\}}_{(j)})\Tr[CU_{I_1}U^T_{I_2}C]\Bigr) =\Tr[F_N^{(4)}\,P^{\text{sign}}]\,,
\end{equation}
where we sum over all matrices and indices thereof in the trace. The constants $c^{\{I\}}_{(j)}$, $d^{\{I\}}_{(j)}$ and $f^{\{I\}}_{(j)}$ come from the combinatorics of the action of the derivatives and the label $\{I\} = \{I_1, I_2\}$. 
Within the single-trace, two-$C$ truncation, the FRGE never generates $\Tr[U_I U_J^T]$ structures:
all projected contributions carry two $C$ insertions.
Consequently, there is no running of $Z_U$ or $Z_V$ at this order, and the flow is entirely driven by the interaction couplings.

We now proceed to the computation of the beta functions $\beta_4^{(j)}$ and $\beta_6^{(j)}$, starting with the latter. In order to obtain contributions to $\beta_6^{(j)}$, we need to look for operators in the expansion of the right-hand side of the FRGE (\ref{eq:FRGEexpansion}) which contain six $U/V$ matrices.
At the order of truncation we are considering for the effective action (equation (\ref{eq:EffectiveActionTruncation}) with $n=3$), contributions can only arise from terms in the form of $F_{N}^{(4)}F_{N}^{(4)}F_{N}^{(4)}$ or $F_{N}^{(4)}F_{N}^{(6)}$.
However, the first combination, once traced with respect to the matrix and sign indices, gives \textit{single trace} terms with the following structure
\begin{multline}
  \sum_{j,k,l}\Bar{g}_4^{(j)}\Bar{g}_4^{(k)}\Bar{g}_4^{(l)}\Big(\sum_{\{I\}}c_{(j,k,l)}^{\{I\}}\Tr[C^3]\Tr[U_{I_1}^TCU_{I_2}U_{I_3}^TCU_{I_4}U_{I_5}^TCU_{I_6}] \\
  +\sum_{\{J\}}d_{(j,k,l)}^{\{J\}}\Tr[\delta]\Tr[CU_{J_1} U_{J_2}^TCCU_{J_2}U_{J_4}^TCCU_{J_5}U_{J_6}^TC]\Big) \\ 
= \Tr[F_{N}^{(4)}P^{\text{sign}}F_{N}^{(4)}P^{\text{sign}}F_{N}^{(4)}P^{\text{sign}}]\,,
\end{multline}
where the constants $c^{\{I\}}_{(j,k,h)}$ and $d^{\{J\}}_{(j,k,h)}$ are combinatorial factors.
In particular, we note that there are no terms proportional to the operator $O_6^{(j)}$ in (\ref{eq:LHSFRGE}) -- the first term vanishes due to $\Tr(C^3)=0$ and the second term has too many $C$s and is therefore projected out.

The second possible contributing term to $\beta_6^{(j)}$ comes from $\Tr[F_{N}^{(4)}P^{\text{sign}}F_{N}^{(6)}P^{\text{sign}}]$ and takes the following form

\begin{multline}
  \sum_{k,l}\Bar{g}_4^{(k)} \Bar{g}_6^{(l)}\Big(\sum_{j}E^{(j)}_{3,(k,l)} \Tr[C^2]O_6^{(j)} 
  + \sum_{\{J\}}c^{\{J\}}_{(k,l)} \Tr[\delta]\Tr[C U_{J_1} U_{J_2}^T C U_{J_3} U_{J_4}^TC U_{J_5} U_{J_6}^TC] \\
+ \sum_{\{K\}} d^{\{K\}}_{(k,l)}\Tr[C]\Tr[U_{K_1}^TCU_{K_2}U_{K_3}^TCU_{K_4}U_{K_5}^TCU_{K_6}]\Big) \\
    =\Tr[F_{N}^{(4)}P^{\text{sign}}F_{N}^{(6)}P^{\text{sign}}],
    \label{eq:F4F6}
\end{multline}
where $(j)$ labels the $U,V$ configuration of the operator associated with each coupling constant $g_6^{(j)}$.
Only the first term matches the structure of the operator $O_6^{(j)}$ in (\ref{eq:LHSFRGE}) and therefore, contributes to $\beta_6^{(j)}$.
The factor $E^{(j)}_{3,(j,k)}$ is the combinatorial factor involved in making the operator $O_6^{(j)}$ in the expansion. From this expression and (\ref{eq:LHSFRGE}), we can read off the form of the beta function equations for $g_6^{(j)}$:
\begin{equation}
  \beta_6^{(j)}=-\alpha_6 g_6^{(j)}+C_1(N)\sum_{k,l}E^{(j)}_{3,(k,l)} g_4^{(k)}g_6^{(l)} \,,
    \label{eq:beta6}
\end{equation}
where we have set the anomalous dimensions to zero as discussed above. For the time being, we put all $N$-dependence in the factor $C_1(N)$ -- we will elaborate on this in the next subsection. 

Let us now compute the beta function for $g_4^{(j)}$. We need terms from the right-hand side of the FRGE that could generate the operators $O_4^{(j)}$ in equation (\ref{eq:LHSFRGE}). These operators require four $U$ or $V$ components, and they can originate from either $F^{(4)}_{N}F^{(4)}_{N}$ or $F^{(6)}_{N}$. Consider the former, which is a second-order contribution.
\begin{multline}
  \sum_{k,l}\Bar{g}_4^{(k)}\Bar{g}_4^{(l)}\Big(\sum_{j}E^{(j)}_{2,(k,l)} \Tr[C^2]O_4^{(j)}
    \\
    + \sum_{\{I\}}f^{\{I\}}_{(k,l)} \Tr[\delta]\Tr[C U_{I_1} U^T_{I_2} C C U_{I_3} U^T_{I_4} C  ]\Big) \\
    = \Tr[F^{(4)}_{N}P^{\text{sign}}F^{(4)}_{N}P^{\text{sign}}].
    \label{eq:F4F4}
\end{multline}
The first term reproduces the operator $O_4^{(j)}$ in (\ref{eq:LHSFRGE}), whereas the second term contains too many $C$s and is projected out. The final contribution comes from the first-order term in the expansion and is given by
\begin{equation}
  \sum_{k}\Bar{g}^{(k)}_6\Big(\sum_{j}D^{(j)}_{2,(k)} \Tr[\delta] O_4^{(j)} +\sum_{\{I\}} r^{\{I\}}_{(k)} \Tr[C]\Tr[U^T_{I_1} C U_{I_2} U^T_{I_3} U_{I_4}]\Big)
  = \Tr[F^{(6)}_{N}P^{\text{sign}}] \,,
\end{equation}
thus contributing to the beta function equations for $g_4^{(j)}$. Any other terms that match the structure of the
operators will be subleading in $N$ and are omitted. In all, the general form of $\beta_4^{(j)}$ is
\begin{equation}
  \beta_4^{(j)}=-\alpha_4 g_4^{(j)}+C_2(N)\sum_{k,l}E^{(j)}_{2,(k,l)}g_4^{(k)}g_4^{(l)}+D(N)\sum_{m}D^{(j)}_{2,(m)}g_6^{(m)} \,, 
    \label{eq:beta4}
\end{equation}
where we have absorbed the $N$-dependence into $C_2(N)$ and $D(N)$.

\subsection{$N$-dependence of the beta functions}
In determining the general form of the beta function equations, we could ignore the factors from $\Dot{R}$ and $P$ in
\eqref{eq:FRGEexpansion} but when it comes to doing calculations, we must treat this more carefully.
Let us consider the term in \eqref{eq:F4F4} that we determined contributes to $\beta_4$. If we include all terms, this expression is given by 
\begin{multline}
  Z_U^{N^{(j)}_U/2}Z_V^{N^{(j)}_V/2}N^{2\alpha_4}\sum_{k,l}E^{(j)}_{2,(k,l)}g_4^{(k)}g_4^{(l)}\\
  \times\sum_{a,b,c,d=1}^N\Dot{R}(a,b) P(a,b)^2 P(c,d)C_{ac}C_{ca}(U_{J_1}^TCU_{J_2})_{bd}(U_{J_3}^TCU_{J_4})_{db}\,,
  \label{eq:F4F4RHS}
\end{multline}
where we have made use of \eqref{eq:CanonicalScalings}. The terms $\Dot{R}(a,b)$ and $P(a,b)$ are given by
\begin{equation}
  \begin{aligned}
    \Dot{R}(a,b) &= \frac{N}{a+b} \theta\Bigl[1-\frac{a+b}{N}\Bigr], \\ 
    P(a,b) &= \Big(\frac{a+b}{N}-1\Big) \theta\Bigl[1-\frac{a+b}{N}\Bigr] + 1,
  \end{aligned}
\end{equation}
Note that the factors of $Z_U$ and $Z_V$ always match up with the appropriate scaling of the operator after using (\ref{eq:CanonicalScalings}) and including the $1/Z_{U/V}$ terms in $P^{\text{sign}}$.
Comparing \eqref{eq:F4F4RHS} to \eqref{eq:LHSFRGE} we have
\begin{multline}
  \sum_{b,d=1}^N(U_{J_1}^TCU_{J_2})_{bd}(U_{J_3}^TCU_{J_4})_{db} \Big(\beta_4^{(j)} + \alpha_4 g_4^{(j)} \\
  - N^{\alpha_4}\sum_{k,l}E^{(j)}_{2,(k,l)}g_4^{(k)}g_4^{(l)}\sum_{a,c=1}^N\Dot{R}(a,b) P(a,b)^2 P(c,d)C_{ac}C_{ca} \Big)=0.
\end{multline}
Since this is true for all $U$ and $V$, then
\begin{equation}
  \beta_4^{(j)} + \alpha_4 g_4^{(j)} - N^{\alpha_4}\sum_{k,l}E^{(j)}_{2,(k,l)}g_4^{(k)}g_4^{(l)}\sum_{a,c=1}^N\Dot{R}(a,b) P(a,b)^2 P(c,d)C_{ac}C_{ca}=0
\end{equation}
for all $b,d$. Reinstating the sum over $b,d$ we get
\begin{equation}
  \beta_4^{(j)} + \alpha_4 g_4^{(j)} - N^{\alpha_4-2}\sum_{k,l}E^{(j)}_{2,(k,l)}g_4^{(k)}g_4^{(l)}\sum_{a,b,c,d=1}^N\Dot{R}(a,b) P(a,b)^2 P(c,d)C_{ac}C_{ca}=0
\end{equation}
Finally, using the fact that the matrix $C$ can be chosen to be diagonal in order to satisfy the constraint \eqref{eq:ConditionC} in the large $N$ limit \cite{Abranches}, we can write the last term as
\begin{equation}
  \sum_{a,b,c,d=1}^N\Dot{R}(a,b) P(a,b)^2 P(c,d)C(a)C(d)\delta_{ac}\delta_{ca}= \sum_{a,b,d=1}^N\Dot{R}(a,b) P(a,b)^2 P(a,d)C(a)^2,
  \label{eq:RPTerm}
\end{equation}
where $C(a)$ are the diagonal terms of $C$. Since the diagonal terms in the large $N$ limit are all bounded from above
\cite{Abranches},
\eqref{eq:RPTerm} scales like $N^3$.
Hence, we can write the beta function equation for $\beta_4^{(j)}$ as 
\begin{equation}
  \beta_4^{(j)}=-\alpha_4 g_4^{(j)}+N^{\alpha_4+1}C_2\sum_{k,l}E^{(j)}_{2,(k,l)} g_4^{(k)}g_4^{(l)} + N^{\alpha_6-\alpha_4+1}D\sum_{m}D^{(j)}_{2,(m)}g_6^{(m)}\,,
  \label{eq:FinalBeta4}
\end{equation}
where $C_2$ depends on the exact calculation of \eqref{eq:RPTerm} and $D$ is found by an analogous calculation coming
from the linear term in \eqref{eq:FRGEexpansion}. 
Carrying out the same calculations for $\beta_6^{(j)}$ we get
\begin{equation}
  \beta_6^{(j)}=-\alpha_6g_6^{(j)}+N^{\alpha_4+1}C_2\sum_{k,l}E^{(j)}_{3,(k,l)} g_4^{(k)}g_6^{(l)}.
  \label{eq:FinalBeta6}
\end{equation}

It is now clear that the upper bounds on the canonical scaling of the couplings are $\alpha_4=-1$ and $\alpha_6 = -2$.
We could now choose to saturate these bounds in order to make the beta functions autonomous, but we allow for the freedom to choose values
below these bounds. This freedom will be useful when analysing the fixed points in the next section.
Note that the constant $C_2$ in \eqref{eq:FinalBeta4} and \eqref{eq:FinalBeta6} is the same. This will be crucial in the
next section when we investigate the fixed points in the regime where $g_6=0$.
The flow depends on three sets of numbers: $E^{(j)}_{2,(k,l)}$ from $F^{(4)}PF^{(4)}$, $E^{(j)}_{3,(k,l)}$ from
$F^{(4)}PF^{(6)}$, and $D^{(j)}_{2,(m)}$ from $F^{(6)}P$. 

\section{Fixed Points Analysis}
\label{sec:Fixed}
Having obtained the beta function equations for the couplings (\ref{eq:beta6}) and (\ref{eq:beta4}), we now analyze the fixed point structure, corresponding to a continuum, large $N$ limit of the matrix model (\ref{eq:IsingCDT}). In particular, we focus on fixed points of the form $(g_4^{*(i)}, 0)$, namely $g_6^{(j)} = 0$. This is a necessary simplification in order to get analytic control of the equations that follow.

Before proceeding, it is useful to set up a convention for the fourth-order operators. This will be useful in the following discussion, where we will refer to couplings and their associated operators interchangeably. The operators corresponding to each $g_4^{(i)}$ coupling are listed in Table \ref{tab:Operators}.
\begin{table}[H]
    \centering
\begin{tabular}{|c|c|}
\hline
    Coupling & Operator \\[0.2cm]
     \hline
    $g_4^{(1)}$   & $Tr[UU^TCUU^TC]$\\[0.2cm]
    \hline
     $g_4^{(2)}$ & $Tr[UU^TCVV^TC]$\\[0.2cm]
     \hline
     $g_4^{(3)}$ &  $Tr[UV^TCUV^TC]$ \\[0.2cm]
     \hline
      $g_4^{(4)}$ & $Tr[UV^TCVU^TC]$ \\[0.1cm]
      \hline
      $g_4^{(5)}$ & $Tr[VV^TCVV^TC]$\\
    \hline
\end{tabular}
\caption{The fourth order operators and their associated couplings.}
    \label{tab:Operators}
\end{table}
If we restrict ourselves to fixed points in the form $(g_4^{*(i)}, 0)$, $\beta_6^{(k)}=0$ is automatically satisfied. The fixed point condition on $\beta_4^{(j)}$ implies
\begin{equation*}
    \beta_4^{(j)}=0 \Rightarrow -\alpha_4 g_4^{*(j)}+C_2(N)\sum_{k,l}E^{(j)}_{2,(k,l)}g_4^{*(k)}g_4^{*(l)}=0 \, .
\end{equation*}
To be completely explicit, we list the non-zero contributions to the beta function equation for $\beta^{(j)}_4$ relevant to the $g^*_6 = 0$ case in Table \ref{tab:E2_nonzero}\footnote{Details of the calculations leading to the results in Table \ref{tab:E2_nonzero} can be found in the accompanying Mathematica notebook. The notebook also computes the coefficients $E_2$ and $D_2$, which are not used in this paper but are included for completeness and to facilitate future work.}.
\begin{table}[H]
\centering
\begin{tabular}{|c|l|}
\hline
$j$ & $(k,l): E^{(j)}_{2,(k,l)}$ \\
\hline
1 & $(1,1):\;16;\quad (4,4):\;4$ \\
2 & $(2,2):\;8;\quad (3,3):\;8$ \\
3 & $(2,3):\;16$ \\
4 & $(1,4):\;16;\quad (4,5):\;16$ \\
5 & $(4,4):\;4;\quad (5,5):\;16$ \\
\hline
\end{tabular}
\caption{Non-zero combinatorial factors $E^{(j)}_{2,(k,l)}$ entering $\beta_4$ via $F^{(4)}PF^{(4)}$ at $g^*_6=0$.}
\label{tab:E2_nonzero}
\end{table}
Hence the beta function equations reduce to 
\[
\begin{aligned}
\beta_{4}^{(1)} &=
-\alpha_{4}\,g_{4}^{(1)}
+ C_{2}(N)\,\Big(16\,(g_{4}^{(1)})^2 + 4\,(g_{4}^{(4)})^2\Big),
\\[4pt]
\beta_{4}^{(2)} &=
-\alpha_{4}\,g_{4}^{(2)}
+ C_{2}(N)\,\Big(8\,(g_{4}^{(2)})^2 + 8\,(g_{4}^{(3)})^2\Big),
\\[4pt]
\beta_{4}^{(3)} &=
-\alpha_{4}\,g_{4}^{(3)}
+ C_{2}(N)\,\Big(16\,g_{4}^{(2)}\,g_{4}^{(3)}\Big),
\\[4pt]
\beta_{4}^{(4)} &=
-\alpha_{4}\,g_{4}^{(4)}
+ C_{2}(N)\,\Big(16\,g_{4}^{(1)}\,g_{4}^{(4)} + 16\,g_{4}^{(4)}\,g_{4}^{(5)}\Big),
\\[4pt]
\beta_{4}^{(5)} &=
-\alpha_{4}\,g_{4}^{(5)}
+ C_{2}(N)\,\Big(4\,(g_{4}^{(4)})^2 + 16\,(g_{4}^{(5)})^2\Big).
\end{aligned}
\]
It is worth noting that the solutions to the above beta-function equations, which we will list momentarily, can be divided into two classes of fixed points, according to their behaviour under the exchange $U \leftrightarrow V$: 
\begin{itemize}
    \item Fixed points which are symmetric under $U \leftrightarrow V$. This condition implies $g_4^{*(1)}=g_4^{*(5)}$;
    \item Fixed points which come in pairs, and are mapped into each other under $U \leftrightarrow V$.
\end{itemize}
Note that this is analogous to what happens in the $ABAB$ matrix model, as pointed out in \cite{Eichhorn2020}.
It is also worth noting that, due to the specific form of the coefficients $E^{(j)}_{2,(k,l)}$, the equations for $g_4^{(1)},g_4^{(4)}, g_4^{(5)}$ completely factorize from those for $g_4^{(2)}, g_4^{(3)}$. Hence, fixed points which live in the five-dimensional parameter space of $g_4^{(i)}$, can be factorized into the Cartesian product of solutions living in a two and a three-dimensional subspace respectively. There are four solutions in the $(g_4^{(2)}, g_4^{(3)})$ subspace, and we summarize them in Table \ref{tab:g2g3}.
\begin{table}[H]
    \centering
\begin{tabular}{|c|cc|}
\hline
    Label & $g_4^{*(2)}$ & $g_4^{*(3)}$\\[0.2cm]
     \hline
    1 &  $0$ & $0$\\[0.2cm]
    2 & $\frac{\alpha_4}{16 C_2(N)}$ & $-\frac{\alpha_4}{16 C_2(N)}$\\[0.2cm]
    3 & $\frac{\alpha_4}{16 C_2(N)}$ &  $\frac{\alpha_4}{16 C_2(N)}$ \\[0.2cm]
    4 &  $\frac{\alpha_4}{8 C_2(N)}$ & 0 \\[0.1cm]
    \hline
\end{tabular}
\caption{Solutions to the beta-function equation in the $(g_4^{(2)}, g_4^{(3)})$ subspace.}
    \label{tab:g2g3}
\end{table}
Note that the corresponding operators are symmetric under $U \leftrightarrow V$, hence the symmetry is preserved at every point in this subspace.

By studying the solutions to the beta-function equation in the $(g_4^{(1)},g_4^{(4)}, g_4^{(5)})$ subspace, we realise that there are two types of solutions. The first type consists of isolated fixed points and will be the object of the next subsection \ref{sec:isol}. The second type is represented by continuous line segments, where every point on the segment is a fixed point of the beta function equations. We will analyse these in subsection \ref{sec:seg}.
\subsection{Isolated fixed points}
\label{sec:isol}
 The isolated fixed points solutions in the $(g_4^{(1)}, g_4^{(4)}, g_4^{(5)})$ subspace are presented in Table \ref{tab:g1g4g5}.
\begin{table}[H]
    \centering
\begin{tabular}{|c|ccc|}
\hline
    Label & $g_4^{*(1)}$ & $g_4^{*(4)}$ & $g_4^{*(5)}$\\[0.2cm]
     \hline
    A &  $0$ & $0$ & $0$\\[0.2cm]
    B & $0$ & $0$ & $\frac{\alpha_4}{16 C_2(N)}$\\[0.2cm]
    C & $\frac{\alpha_4}{16 C_2(N)}$ &  $0$ & $0$\\[0.2cm]
    D &  $\frac{\alpha_4}{16 C_2(N)}$ & 0 & $\frac{\alpha_4}{16 C_2(N)}$\\[0.1cm]
    \hline
\end{tabular}
\caption{Solutions to the beta-function equation in the $(g_4^{(1)}, g_4^{(4)}, g_4^{(5)})$ subspace.}
    \label{tab:g1g4g5}
\end{table}
Here, solutions A and D are symmetric under $U \leftrightarrow V$, whereas solutions B and C are mapped one into the other by the same transformation. Note that we have chosen a labelling that allows us to refer to the solution in the full space by combining solutions in the two subspaces. For example, solution 2B corresponds to 
\begin{equation*}
    (g_4^{*(1)},g_4^{*(2)},g_4^{*(3)},g_4^{*(4)},g_4^{*(5)})_{2B}=\Bigl(0,\frac{\alpha_4}{16 C_2(N)},-\frac{\alpha_4}{16 C_2(N)}, 0, \frac{\alpha_4}{16 C_2(N)}\Bigr) \,.
\end{equation*}
Therefore, there are $16$ solutions in total, corresponding to all pairings across the two tables. Note that in all solutions, $g_4^{*(4)}=0$. This is analogous to what happens in the $ABAB$ matrix model \cite{Eichhorn2020}, where the coupling associated with the operator $Tr[ABBA]$ always vanishes at the fixed point.

Among these fixed points, there are some which lend themselves to a straightforward interpretation:
\begin{itemize}
    \item Gaussian fixed point (solution 1A)
    \begin{equation}
       g_4^{*(i)}=(0,0,0,0,0) \, .
       \label{eq:GaussianFixed}
    \end{equation}
    On this fixed point, we have a free field theory;
    \item Pure CDT fixed points (solutions 1B and 1C)
    \begin{align}
    \label{eq:pureCDT1}
         g_4^{*(i)}&=\Bigl(\frac{\alpha_4}{16 C_2(N)}, 0, 0, 0, 0\Bigr) \\
          g_4^{*(i)}&=\Bigl(0, 0, 0, 0, \frac{\alpha_4}{16 C_2(N)}\Bigr)\, .
    \label{eq:pureCDT2}
    \end{align}
    On these fixed points, there is only one independent, non-vanishing coupling, representing the cosmological constant, which couples to the area operator. A fixed point of this kind also occurs in the analysis of the RG flow for the matrix model of pure CDT \cite{castro2020renormalization}. Therefore, we interpret this as a continuum limit in which the Ising model is switched off, and we are left with a theory of pure gravity.
\end{itemize}
Other solutions to the fixed-point equations represent non-trivial fixed points, which are specific to the matrix model we are considering.

The next step, in order to characterize these fixed points, is to compute the critical exponents, defined as the eigenvalues of the Hessian matrix $H_{ij}=-\frac{\partial \beta_i}{\partial g_j}$. They determine how couplings grow as we flow away from the critical point. In particular, a positive critical exponent corresponds to a relevant direction, a negative critical exponent characterizes an irrelevant direction and a critical exponent of $0$ is called marginal. Taking derivatives of (\ref{eq:beta6}) and (\ref{eq:beta4}), the Hessian matrix takes the block form 
\[
    H=\begin{pNiceArray}{c|c}
        \alpha_4 \delta^i_j-C_2(N)E^{(i)}_{2,(j,k)}g_4^{*(k)} & D(N)D^{(i)}_{2,(l)} \\
        \hline
        C_2(N)E^{(h)}_{3,(j,k)} g_6^{*(k)} & \alpha_6 \delta^h_l-C_2(N)E^{(h)}_{3,(l,k)} g_4^{*(k)}
    \end{pNiceArray} \, ,
      \]  
    with $i, j=1,..,5$ and $h,l=1,...,20$. On the fixed points we are considering, this simplifies to 
    \[
    H=\begin{pNiceArray}{c|c}
        \alpha_4 \delta^i_j-C_2(N)E^{(i)}_{2,(j,k)}g_4^{*(k)} & D(N)D^{(i)}_{2,(j)} \\
        \hline
        \Block{1-1}<\Large>{0} & \alpha_6 \delta^h_l-C_2(N)E^{(h)}_{3,(l,k)} g_4^{*(k)}
    \end{pNiceArray} \, ,
      \]  
so that the overall eigenvalues are given by the eigenvalues of the two blocks on the diagonal separately. 
Firstly, let us consider the critical exponents corresponding to the pure CDT solutions: 
\begin{itemize}
    \item Gaussian fixed point (solution 1A). The critical exponents corresponding to the upper-left block of the Hessian are
    \begin{equation*}
        \lambda_4^{(i)}=\alpha_4, \quad \forall i=1,...,5 \, ,
    \end{equation*}
    and the eigenvalues of the bottom-right block are given by
    \begin{equation*}
        \lambda_6^{(h)}=\alpha_6, \quad \forall h=1,...,20 \, .
    \end{equation*}
    All directions of the RG flow from the Gaussian fixed point are irrelevant, and the theory is free, given $\alpha_4, \alpha_6<0$. The conditions on $\alpha_4$ and $\alpha_6$ are satisfied when the bounds are saturated. This continuum limit corresponds to topological gravity.
    \item Pure CDT fixed points. For both solutions 1B (\ref{eq:pureCDT1}) and 1C (\ref{eq:pureCDT2}), the critical exponents in the $g_4^{(i)}$ subspace are 
    \begin{equation*}
        \lambda_4^{(i)}=(\alpha_4, \alpha_4, 0, -\alpha_4, \alpha_4) \, ,
    \end{equation*}
    therefore there are three irrelevant, one marginal and one relevant direction. On the bottom-right block, critical exponents associated with this solution are 
    \begin{align*}
        \lambda_6^{(h)}&=\alpha_6, \,\quad  \text{for}\quad  h=1,...,12 \\ \lambda_6^{(h)}&=-\alpha_4+\alpha_6, \,\quad  \text{for} \quad h=13,...,18 \\ \lambda_6^{(h)}&=-2\alpha_4+\alpha_6, \, \quad \text{for}\quad h=19,20\, .
    \end{align*} 
    It follows that, for $\frac{\alpha_6}{\alpha_4}>2$, all those directions are irrelevant. In particular, for the canonical values of $\alpha_4$ and $\alpha_6$, 18 directions are irrelevant and two are marginal. Hence, within this region of parameter space, there is only one relevant direction. This is consistent with the result of \cite{castro2020renormalization}, for the continuum limit of the matrix model describing pure CDT, with the same value of the critical exponent for the relevant direction. Hence, this fixed point corresponds to gravity with a cosmological constant term and no matter. It was shown in \cite{castro2020renormalization} that, for the model of pure CDT, there is another fixed point which corresponds to the continuum limit of 2D gravity with a cosmological constant for complementary values of $\alpha_4$ and $\alpha_6$. We believe that such a fixed point should also be present in the Ising--CDT model. However, it will only exist in the $g^{(i)}_6 \neq 0$ regime, where a full analytic or numerical study would be required to prove its existence.
    \end{itemize}
    Critical exponents corresponding to all solutions of the fixed-point equations are presented in Appendix \ref{app:Crit}. One of the main features of solutions characterizing non-trivial fixed points is that critical exponents in the bottom-right block only take three values, with different degeneracies depending on the solution.  Those values are $\alpha_6$, $\alpha_6-\alpha_4$ and $\alpha_6-2\alpha_4$. This implies that those directions are all irrelevant or marginal for the canonical values of $\alpha_4, \alpha_6$, and they are all irrelevant in the region $\frac{\alpha_6}{\alpha_4}>2$. This means that, as far as this region is concerned, there are no relevant directions coming from order six operators. Indeed, in different regions of parameter space, some of those directions will become relevant.

Let us now focus on the upper-left block. Different solutions are characterized by a different number of relevant, irrelevant and marginal directions. Note that critical exponents take three values $\alpha_4$, $0$ and $-\alpha_4$, corresponding to irrelevant, marginal and relevant directions respectively. In particular:
\begin{itemize}
    \item Solution 1D has 2 irrelevant and 3 relevant directions;
    \item Solutions 2A and 3A have 4 irrelevant and 1 relevant directions; 
    \item Solutions 2B, 3B, 2C and 3C have 2 irrelevant, 1 marginal and 2 relevant directions;
    \item Solutions 2D and 3D have 1 irrelevant and 4 relevant directions;
    \item Solution 4A has 3 irrelevant and 2 relevant directions;
    \item Solutions 4B and 4C have 1 irrelevant, 1 marginal and 3 relevant directions;
    \item Solution 4D has 5 relevant directions.
\end{itemize}
To summarize, besides recovering the fixed points corresponding to topological gravity and pure gravity with a cosmological constant, which can be obtained as the continuum limit of the pure CDT and no matter matrix model \cite{castro2020renormalization}, we found new, non-trivial fixed points. In particular, within the region of the $\alpha_4$, $\alpha_6$ parameter space specified above, three of those fixed points, namely 1D, 4B and 4C, have three relevant directions. It follows that the corresponding CFT has the correct number of operators to represent the Ising CFT coupled to two-dimensional gravity. The operators in the Ising CFT are, namely, the cosmological constant term (identity), the energy operator $\psi$, and the duality operator $\sigma$.  
\subsection{Segments of fixed points}
\label{sec:seg}
\begin{figure}
  \begin{center}
    \includegraphics[width=0.80\textwidth]{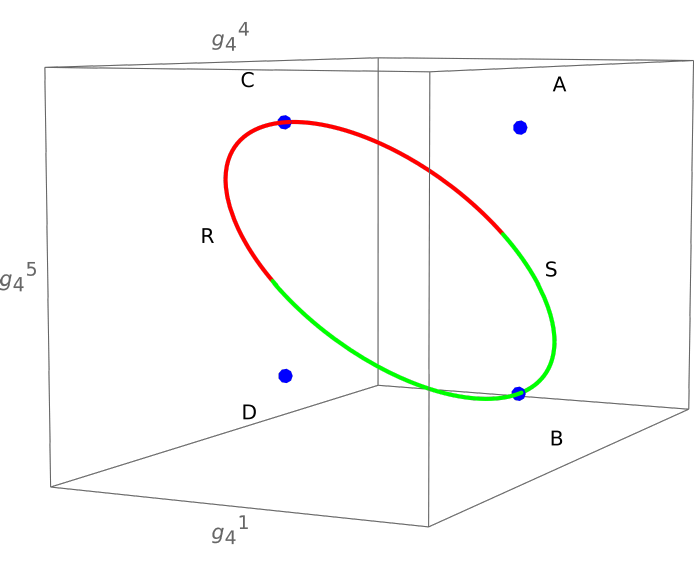}
  \end{center}
  \caption{Projection of the phase space of the theory to the $(g_4^{(1)}, g_4^{(4)}, g_4^{(5)})$ subspace. This is composed by solutions A, B, C, D (blue), and critical segments R (red) and S (green).}\label{fig:phase145}
\end{figure}

Alongside isolated fixed point solutions, beta-function equations in the $(g_4^{(1)},g_4^{(4)}, g_4^{(5)})$ subspace admit lines of solutions such that each point on the line is a fixed point. We present them in Table \ref{tab:segment}.
\begin{table}[H]
    \centering
\begin{tabular}{|c|ccc|}
\hline
    Label & $g_4^{*(1)}$ & $g_4^{*(4)}$ & $g_4^{*(5)}$\\[0.2cm]
     \hline
    R &  $\frac{1}{32C_2(N)}\bigl(\alpha_4-s(g_4^{*(4)};\alpha_4)\bigr)$ & $\in \Bigl(-\frac{|\alpha_4|}{16C_2(N)},\frac{|\alpha_4|}{16C_2(N)}\Bigr)$ & $\frac{1}{32C_2(N)}\bigl(\alpha_4+s(g_4^{*(4)};\alpha_4)\bigr)$\\[0.5cm]
    S & $\frac{1}{32C_2(N)}\bigl(\alpha_4+s(g_4^{*(4)};\alpha_4)\bigr)$ &  $\in \Bigl(-\frac{|\alpha_4|}{16C_2(N)},\frac{|\alpha_4|}{16C_2(N)}\Bigr)$ & $\frac{1}{32C_2(N)}\bigl(\alpha_4-s(g_4^{*(4)};\alpha_4)\bigr)$\\[0.1cm]
    \hline
\end{tabular}
\caption{Lines of solutions to the beta-function equation in the $(g_4^{(1)}, g_4^{(4)}, g_4^{(5)})$ subspace. Here $s(g_4^{*(4)};\alpha_4)=\sqrt{\alpha_4^2-256g_4^{*(4)\, 2}C_2(N)^2}$.}
    \label{tab:segment}
\end{table}
The solutions in the extended parameter space are obtained by crossing entries of Table \ref{tab:segment} with the ones in Table \ref{tab:g2g3}. It is worth noting that fixed points of type B and C from table \ref{tab:g1g4g5} can be obtained as particular points of the segments R and S respectively, with $g_4^{*(4)}=0$. It follows that solutions of type B and C sit on the segments. These include the pure CDT solutions 1B (\ref{eq:pureCDT1}) and 1C (\ref{eq:pureCDT2}), which sit on segments 1R and 1S respectively. We show, in Figure \ref{fig:phase145}, the projection of the phase space (restricted to the $(g_4^{*(i)}, 0)$ condition) to the $(g_4^{(1)}, g_4^{(4)}, g_4^{(5)})$ subspace, which includes critical points from Table \ref{tab:g1g4g5} and critical segments from Table \ref{tab:segment}.

Along each segment one finds \(g_4^{(1)}+g_4^{(5)}=\alpha_4/(16\,C_2)\) while \(g_4^{(4)}\) varies within the interval in Table~\ref{tab:segment}; we verified that (i) the spectrum of critical exponents and (ii) the autonomous beta polynomials are independent of \(g_4^{(4)}\) along the segment. This indicates that the segment direction is generated by a redundant (reparameterization) vector field in theory space: within the single-trace, two-\(C\) basis, a linear redefinition among \(\{O_4^{(1)},O_4^{(4)},O_4^{(5)}\}\) leaves projected flows and spectra invariant, so points on the same segment are physically equivalent at the level of this truncation. A decisive test would require either (a) adding interactions outside this basis (e.g.\ multi-trace or three-\(C\) operators), or (b) computing an observable sensitive to \(g_4^{(4)}\) (for instance a suitable four-point function from \(\Gamma_N\)) and checking its constancy along the segment. We leave this for future work. Consistent with this picture, Appendix~\ref{app:Crit}, Table~\ref{tab:degeneracysegment}, shows that every point on a given segment has identical critical exponents equal to those of the corresponding B or C fixed point with exactly one marginal direction in the quartic sector, interpreted as the RG flow along the segment. In particular, solutions 4R and 4S display three relevant directions and are therefore suitable to represent the Ising CFT.

\section{Conclusion}
In this work, we investigated the Ising model coupled to Causal Dynamical Triangulations (CDT) through the lens of the Functional Renormalization Group Equation (FRGE). Using a matrix model representation, we analyzed the fixed-point structure of the theory and demonstrated the existence of continuum limits. Our results confirm that the model not only reproduces known fixed points, such as those corresponding to topological gravity and pure CDT with a cosmological constant but also uncovers a new fixed point featuring three relevant directions, aligning with the number of primary operators of the Ising CFT.

The addition of Ising spin degrees of freedom significantly increased the complexity of the truncation, making the fixed-point analysis more challenging. However, by leveraging symmetry and simplifying assumptions, we successfully identified physically relevant fixed points, providing further evidence that the Ising CFT can emerge in this setting.

Future research could focus on extending this analysis to complementary regions of the parameter space (i.e. $g_6 \neq 0$) and exploring fixed points beyond the analytic reach of the current truncation. Numerical methods, in particular, could shed light on the existence of additional fixed points and their critical properties, providing a more comprehensive picture of the Ising-CDT model. The analysis could be further improved by adding higher order terms in the effective action, and by extending the truncation to also include multi-trace operators or operators with more than two C matrices. This would provide more quantitative results for the critical exponents.

\acknowledgments

We would like to thank John Wheater for reviewing and commenting on this manuscript. Ryan Barouki is supported by STFC studentship ST/W507726/1 and Davide Laurenzano is supported by STFC studentship ST/X508664/1.

\appendix
\section{Critical exponents for the RG fixed points}
\label{app:Crit}
In this section, we present the values of the critical exponents, obtained by diagonalizing the Hessian matrix $H_{i j}=-\frac{\partial\beta_i}{\partial g_j}$. As explained in the main text, eigenvalues in the upper-left block take three different values, corresponding to irrelevant, marginal and relevant directions. We specify the degeneracy of these eigenvalues for each solution of the beta-function equations in the left half of Table \ref{tab:expg4}. We have also seen that in the bottom-right block of the Hessian, the eigenvalues take three possible values. We specify the degeneracies of these eigenvalues for each solution on the right half of the same table.
\begin{table}[H]
    \centering
    \begin{tabular}{c|ccc|ccc}
    \hline
        Solution & $\alpha_4$ & 0 & -$\alpha_4$ & $\alpha_6$ & $\alpha_6-\alpha_4$ & $\alpha_6-2\alpha_4$\\[0.2cm]
        \hline 
         1A & 5& 0&0 & 20& 0&0\\[0.2cm]
         1B & 3& 1&1 & 12& 6&2\\[0.2cm]
         1C & 3 & 1&1& 12 & 6&2\\[0.2cm]
         1D & 2 & 0 & 3 & 6 & 8 & 6\\[0.2cm]
         2A & 4 & 0 & 1 & 12 & 6 & 2\\[0.2cm]
         2B & 2 & 1 & 2 & 6 & 8 & 6\\[0.2cm]
         2C & 2 & 1 & 2 & 6 & 8 & 6\\[0.2cm]
         2D & 1 & 0 & 4 & 2 & 6 & 12\\[0.2cm]
         3A & 4 & 0 & 1 & 12 & 6 & 2\\[0.2cm]
         3B & 2 & 1 & 2 & 6 & 8 & 6\\[0.2cm]
         3C & 2 & 1 & 2 & 6 & 8 & 6\\[0.2cm]
         3D & 1 & 0 & 4 & 2 & 6 & 12\\[0.2cm]
         4A & 3 & 0 & 2 & 6 & 8 & 6\\[0.2cm]
         4B & 1 & 1 & 3 & 2 & 6 & 12\\[0.2cm]
         4C & 1 & 1 & 3 & 2 & 6 & 12\\[0.2cm]
         4D & 0 & 0 & 5 & 0 & 0 & 20\\[0.2cm]
         \hline
    \end{tabular}
    \caption{Degeneracy of the eigenvalues in both the upper-left and the bottom-right blocks of the Hessian for each
    fixed point.}
    \label{tab:expg4}
\end{table}
Note that the order six operators are all irrelevant for $\frac{\alpha_6}{\alpha_4}>2$ and all irrelevant or marginal for the canonical values $\alpha_4=-1$ and $\alpha_6=-2$. 

We list the degeneracies of the critical exponents on the fixed line segments (section \ref{sec:seg}) in Table \ref{tab:degeneracysegment}.
\begin{table}[H]
    \centering
    \begin{tabular}{c|ccc|ccc}
    \hline
        Solution & $\alpha_4$ & 0 & -$\alpha_4$ & $\alpha_6$ & $\alpha_6-\alpha_4$ & $\alpha_6-2\alpha_4$\\[0.2cm]
        \hline 
         1R & 3& 1&1 & 12& 6&2\\[0.2cm]
         1S & 3 & 1&1& 12 & 6&2\\[0.2cm]
         2R & 2 & 1 & 2 & 6 & 8 & 6\\[0.2cm]
         2S & 2 & 1 & 2 & 6 & 8 & 6\\[0.2cm]
         3R & 2 & 1 & 2 & 6 & 8 & 6\\[0.2cm]
         3S & 2 & 1 & 2 & 6 & 8 & 6\\[0.2cm]
         4R & 1 & 1 & 3 & 2 & 6 & 12\\[0.2cm]
         4S & 1 & 1 & 3 & 2 & 6 & 12\\[0.2cm]
         \hline
    \end{tabular}
    \caption{Degeneracy of the eigenvalues in both the upper-left and the bottom-right blocks of the Hessian for fixed line
    segments.}
    \label{tab:degeneracysegment}
\end{table}
Segments of fixed points have the same critical exponents as the corresponding B and C type solutions. They feature exactly one marginal direction in the order four parameter space, corresponding to the direction along the segment.
\bibliographystyle{JHEP}
\bibliography{references.bib}

\providecommand{\href}[2]{#2}\begingroup\raggedright\begin{thebibliography}{10}

\bibitem{Regge1961}
T.~Regge, \emph{{General Relativity Without Coordinates}}, \href{https://doi.org/10.1007/BF02733251}{\emph{Nuovo Cim.} {\bfseries 19} (1961) 558}.

\bibitem{Thooft1974}
G.~Hooft, \emph{A planar diagram theory for strong interactions}, \href{https://doi.org/https://doi.org/10.1016/0550-3213(74)90154-0}{\emph{Nuclear Physics B} {\bfseries 72} (1974) 461}.

\bibitem{Saad2019}
P.~Saad, S.H.~Shenker and D.~Stanford, \emph{{JT gravity as a matrix integral}},  \href{https://arxiv.org/abs/1903.11115}{{\ttfamily 1903.11115}}.

\bibitem{Collier}
S.~Collier, L.~Eberhardt, B.~Muehlmann and V.A.~Rodriguez, \emph{{The Virasoro minimal string}}, \href{https://doi.org/10.21468/SciPostPhys.16.2.057}{\emph{SciPost Phys.} {\bfseries 16} (2024) 057} [\href{https://arxiv.org/abs/2309.10846}{{\ttfamily 2309.10846}}].

\bibitem{Collier2024}
S.~Collier, L.~Eberhardt, B.~M\"uhlmann and V.A.~Rodriguez, \emph{{The complex Liouville string: the matrix integral}},  \href{https://arxiv.org/abs/2410.07345}{{\ttfamily 2410.07345}}.

\bibitem{DiFrancesco1993}
P.~Di~Francesco, P.H.~Ginsparg and J.~Zinn-Justin, \emph{{2-D Gravity and random matrices}}, \href{https://doi.org/10.1016/0370-1573(94)00084-G}{\emph{Phys. Rept.} {\bfseries 254} (1995) 1} [\href{https://arxiv.org/abs/hep-th/9306153}{{\ttfamily hep-th/9306153}}].

\bibitem{Anninos2020}
D.~Anninos and B.~M\"uhlmann, \emph{{Notes on matrix models (matrix musings)}}, \href{https://doi.org/10.1088/1742-5468/aba499}{\emph{J. Stat. Mech.} {\bfseries 2008} (2020) 083109} [\href{https://arxiv.org/abs/2004.01171}{{\ttfamily 2004.01171}}].

\bibitem{Ambjorn1998}
J.~Ambjorn and R.~Loll, \emph{{Nonperturbative Lorentzian quantum gravity, causality and topology change}}, \href{https://doi.org/10.1016/S0550-3213(98)00692-0}{\emph{Nucl. Phys. B} {\bfseries 536} (1998) 407} [\href{https://arxiv.org/abs/hep-th/9805108}{{\ttfamily hep-th/9805108}}].

\bibitem{Durhuus2009sm}
B.~Durhuus, T.~Jonsson and J.F.~Wheater, \emph{{On the spectral dimension of causal triangulations}}, \href{https://doi.org/10.1007/s10955-010-9968-x}{\emph{J. Statist. Phys.} {\bfseries 139} (2010) 859} [\href{https://arxiv.org/abs/0908.3643}{{\ttfamily 0908.3643}}].

\bibitem{Ambj_rn_2006}
J.~Ambj{\o}rn, J.~Jurkiewicz and R.~Loll, \emph{The universe from scratch}, \href{https://doi.org/10.1080/00107510600603344}{\emph{Contemporary Physics} {\bfseries 47} (2006) 103}.

\bibitem{Ambj_rn_1999}
J.~Ambj{\o}rn, K.N.~Anagnostopoulos and R.~Loll, \emph{New perspective on matter coupling in 2d quantum gravity}, \href{https://doi.org/10.1103/physrevd.60.104035}{\emph{Physical Review D} {\bfseries 60} (1999) }.

\bibitem{Ambj_rn_2000}
J.~Ambj{\o}rn, K.~Anagnostopoulos and R.~Loll, \emph{Crossing the c=1 barrier in 2d lorentzian quantum gravity}, {\emph{Physical Review D} {\bfseries 61} (2000) 44010}.

\bibitem{Ambj_rn_2009}
J.~Ambj{\o}rn, K.~Anagnostopoulos, R.~Loll and I.~Pushkina, \emph{Shaken, but not stirred{\textemdash}potts model coupled to quantum gravity}, \href{https://doi.org/10.1016/j.nuclphysb.2008.08.030}{\emph{Nuclear Physics B} {\bfseries 807} (2009) 251}.

\bibitem{ambjorn_c_2015}
J.~Ambjørn, A.~Görlich, J.~Jurkiewicz and H.~Zhang, \emph{A \textit{c} = 1 phase transition in two-dimensional {CDT}/{Horava}–{ Lifshitz} gravity?}, \href{https://doi.org/10.1016/j.physletb.2015.03.008}{\emph{Physics Letters B} {\bfseries 743} (2015) 435}.

\bibitem{Ambj_rn_2013}
J.~Ambjørn, L.~Glaser, Y.~Sato and Y.~Watabiki, \emph{2d cdt is 2d hořava–lifshitz quantum gravity}, \href{https://doi.org/10.1016/j.physletb.2013.04.006}{\emph{Physics Letters B} {\bfseries 722} (2013) 172–175}.

\bibitem{Wheater_2022}
J.F.~Wheater and P.D.~Xavier, \emph{The cylinder amplitude in the hard dimer model on 2d causal dynamical triangulations}, \href{https://doi.org/10.1088/1361-6382/ac50ec}{\emph{Classical and Quantum Gravity} {\bfseries 39} (2022) 075004}.

\bibitem{durhuus2022trees}
B.~Durhuus, T.~Jonsson and J.~Wheater, \emph{From trees to gravity},  \href{https://arxiv.org/abs/2211.15247}{{\ttfamily 2211.15247}}.

\bibitem{Barouki2025}
R.~Barouki, H.~Stubbs and J.~Wheater, \emph{Conformal dimensions on causal random geometry},  \href{https://arxiv.org/abs/2501.17930}{{\ttfamily 2501.17930}}.

\bibitem{Ambjorn2005}
J.~Ambjorn, J.~Jurkiewicz and R.~Loll, \emph{{Reconstructing the universe}}, \href{https://doi.org/10.1103/PhysRevD.72.064014}{\emph{Phys. Rev. D} {\bfseries 72} (2005) 064014} [\href{https://arxiv.org/abs/hep-th/0505154}{{\ttfamily hep-th/0505154}}].

\bibitem{Loll2005}
J.~Ambjorn, J.~Jurkiewicz and R.~Loll, \emph{{Spectral dimension of the universe}}, \href{https://doi.org/10.1103/PhysRevLett.95.171301}{\emph{Phys. Rev. Lett.} {\bfseries 95} (2005) 171301} [\href{https://arxiv.org/abs/hep-th/0505113}{{\ttfamily hep-th/0505113}}].

\bibitem{Ambjorn2007}
J.~Ambjorn, A.~Gorlich, J.~Jurkiewicz and R.~Loll, \emph{{Planckian Birth of the Quantum de Sitter Universe}}, \href{https://doi.org/10.1103/PhysRevLett.100.091304}{\emph{Phys. Rev. Lett.} {\bfseries 100} (2008) 091304} [\href{https://arxiv.org/abs/0712.2485}{{\ttfamily 0712.2485}}].

\bibitem{Abranches}
J.L.A.~Abranches, A.D.~Pereira and R.~Toriumi, \emph{{Dually weighted multi-matrix models as a path to causal gravity-matter systems}},  \href{https://arxiv.org/abs/2310.13503}{{\ttfamily 2310.13503}}.

\bibitem{Eichhorn2013}
A.~Eichhorn and T.~Koslowski, \emph{{Continuum limit in matrix models for quantum gravity from the Functional Renormalization Group}}, \href{https://doi.org/10.1103/PhysRevD.88.084016}{\emph{Phys. Rev. D} {\bfseries 88} (2013) 084016} [\href{https://arxiv.org/abs/1309.1690}{{\ttfamily 1309.1690}}].

\bibitem{castro2020renormalization}
A.~Castro and T.~Koslowski, \emph{Renormalization group approach to the continuum limit of matrix models of quantum gravity with preferred foliation}, \href{https://doi.org/10.3389/fphy.2021.531766}{\emph{Frontiers in Physics} {\bfseries 9} (2021) } [\href{https://arxiv.org/abs/2008.10090}{{\ttfamily 2008.10090}}].

\bibitem{Benedetti}
D.~Benedetti and J.~Henson, \emph{{Imposing causality on a matrix model}}, \href{https://doi.org/10.1016/j.physletb.2009.06.027}{\emph{Phys. Lett. B} {\bfseries 678} (2009) 222} [\href{https://arxiv.org/abs/0812.4261}{{\ttfamily 0812.4261}}].

\bibitem{Boulatov1986}
D.V.~Boulatov and V.A.~Kazakov, \emph{{The Ising Model on Random Planar Lattice: The Structure of Phase Transition and the Exact Critical Exponents}}, \href{https://doi.org/10.1016/0370-2693(87)90312-1}{\emph{Phys. Lett. B} {\bfseries 186} (1987) 379}.

\bibitem{Wetterich_1993}
C.~Wetterich, \emph{Exact evolution equation for the effective potential}, \href{https://doi.org/10.1016/0370-2693(93)90726-x}{\emph{Physics Letters B} {\bfseries 301} (1993) 90–94} [\href{https://arxiv.org/abs/1710.05815}{{\ttfamily 1710.05815}}].

\bibitem{Eichhorn2018}
A.~Eichhorn, T.~Koslowski and A.D.~Pereira, \emph{Status of background-independent coarse-graining in tensor models for quantum gravity},  \href{https://arxiv.org/abs/1811.12909}{{\ttfamily 1811.12909}}.

\bibitem{Eichhorn2020}
A.~Eichhorn, A.D.~Pereira and A.G.A.~Pithis, \emph{{The phase diagram of the multi-matrix model with ABAB-interaction from functional renormalization}}, \href{https://doi.org/10.1007/JHEP12(2020)131}{\emph{JHEP} {\bfseries 12} (2020) 131} [\href{https://arxiv.org/abs/2009.05111}{{\ttfamily 2009.05111}}].

\end{thebibliography}\endgroup
\end{document}